%% file: survey.tex
\documentclass[10pt,conference]{IEEEtran}

\usepackage{cite}
\usepackage{comment}
\usepackage{xspace}
\usepackage{multirow}
\usepackage{url}
\usepackage{hyperref}

\usepackage{float}
\usepackage{tikz}
\usepackage{amsmath,amssymb,amsfonts}
\usepackage{algorithm}
\usepackage{algorithmic}
\usepackage{comment}
\usepackage{graphicx}
\usepackage[numbers,compress]{natbib}
\usepackage{soul}
\usepackage{xcolor}
\usepackage{multirow}

\usepackage{adjustbox}

\newcommand{\ie}{{\itshape i.e.}\xspace}
\newcommand{\eg}{\emph{e.g.}\xspace}
\newcommand{\etc}{\emph{etc.}\xspace}

\makeatletter

\begin{document}

\newcommand*\circled[1]{\tikz[baseline=(char.base)]{
            \node[shape=circle,fill,inner sep=0pt] (char) {\textcolor{white}{#1}};}}
            
\title{A Survey of \textit{Multi-Tenant}\\Deep Learning Inference on GPU}

\author{Fuxun Yu{$^\dagger$}, Di Wang{$^\ddagger$}, Longfei Shangguan{$^\ddagger$}, Minjia Zhang{$^\ddagger$}, Chenchen Liu$^\S$, Tolga Soyata{$^\dagger$}, Xiang Chen{$^\dagger$}\\
{{$^\dagger$}George Mason University, {$^\ddagger$}Microsoft, {$^\S$}University of Maryland, Baltimore County}
}


\maketitle

\input{_txt/0_abs}

\input{_txt/1_intro}
\input{_txt/2_def}

\input{_txt/3_tech}

\input{_txt/4_future}
\input{_txt/6_conclusion}

\small
\bibliographystyle{ACM-Reference-Format}
\bibliography{sample-base}

\end{document}

%% file: _txt/0_abs.tex
\begin{abstract}
  Deep Learning (DL) models have achieved superior performance.
  Meanwhile, the computing hardware like NVIDIA GPUs also demonstrated strong computing scaling trends with 2$\times$ throughput and memory bandwidth for each generation.
  With such strong computing scaling of GPUs, \textit{multi-tenant deep learning inference} by co-locating multiple DL models onto the same GPU become widely deployed to improve resource utilization, enhance serving throughput, and reduce energy cost, etc.
However, achieving efficient multi-tenant DL inference is challenging which requires thorough full-stack system optimization.
Previous surveys either target at summarizing single tenant deep learning inference optimizations, or only focus on certain single optimization layer alone, such as graph-level, kernel-level, etc.
  This survey aims to summarize and categorize the emerging challenges and optimization opportunities for multi-tenant DL inference on GPU. 
  By overviewing the entire optimization stack, summarizing the multi-tenant computing innovations, and elaborating the recent technique advances, we hope that this survey could shed light on new optimization perspectives and motivate novel works in future large-scale DL inference system optimization.  
\end{abstract}


%% file: _txt/1_intro.tex
\section{\textbf{Introduction}}


	
\begin{figure*}[!tb]
  \centering
  \vspace{-2mm}
  \includegraphics[width=6.9in]{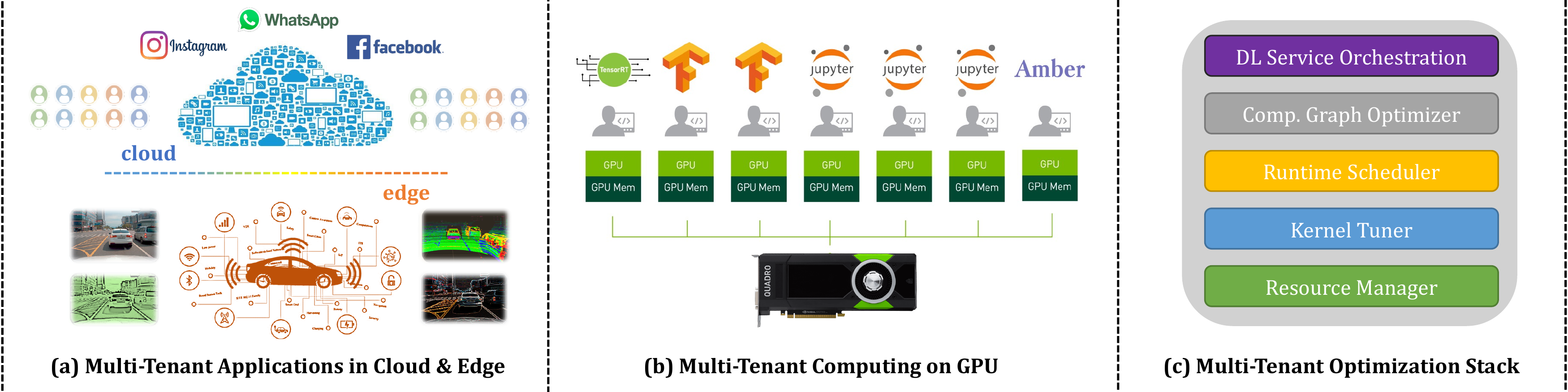}
  \vspace{-2mm}
  \caption{The Emerging Trend of Multi-Tenant DL Computing on GPU.}
  \label{fig:1}
  \vspace{-4mm}
\end{figure*}

\textbf{DL Application and Computing Trends} Deep Learning (DL) models have achieved superior performance in cognitive tasks like vision, speech and language domain, and have been adopted in medical analysis, machine translation, product recommendation, \etc 
	The momentum of DL-based intelligence has appealed millions of users and created a wide-spectrum of cloud \& edge applications like VR/AR games, intelligent robots and vehicles, large-scale recommendation systems, and even metaverse applications~\cite{metaverse}, many of which are featured with \textit{multi-modality, multi-tasking and substantial task complexity}, as shown in Figure~\ref{fig:1} (a).

The emergence of such massive DL applications motivates the adoption of DL accelerators,
especially GPUs, in both cloud and edge.
According to the report~\cite{gpu_report}, GPUs accounted for 85\% of the \$2.98B cloud data center accelerator market in 2018.
The edge hardware market, with the emerging smart manufacturing, surveillance applications, is also projected to grow from \$920M in 2021 to \$2,080M by 2026 and the edge GPUs are also taking a steady growth to more than 50\% market share with Nvidia Jetson, TX2, Xavier, Orin, etc.

Within such trends, the capacity of recent generations of GPUs demonstrates exponential growing speed.\footnote{
	Detailed scaling statistics of GPU capacities could be found in \cite{gpu_stat}.}
	From K80, P40, and P100 to recent T40, V100, A100 architectures, GPUs maintain a trend of doubling performance.
	The last generation of V100~\cite{v100} offers 120 Tera floating point operations per second (TFLOPS) and 900 GB/s memory bandwidth, and the numbers further increase to 312 TFLOPS and 1.6TB/s memory bandwidth for the newer A100~\cite{a100}. A100 reports the ResNet50~\cite{r50} inference speed of 36,436 images/second, showing the computing capacity that overwhelms the limited needs from conventional single DL model execution schemes.
Therefore, with such scaling trends in both application complexity and GPU capacity, single model execution cannot fulfill the needs of application scenarios nor fully utilizing the GPUs.

\vspace{0.5mm}
\textbf{Multi-Tenant DL Inference}, as shown in Figure~\ref{fig:1} (b), is \textit{one promising solution to the aforementioned scenarios with multi-modality and multi-tasking needs by running mixed DL model workloads simultaneously on one powerful GPU to improve the utilization, throughput, and power efficiency, etc.} 

There are recently many emerging works that tackle the multi-tenant DL inference optimization on GPUs. These works usually take single optimization point of view drawing from traditional single-model optimization experiences, e.g., either from the DL model scheduling perspective~\cite{iccad,ios,auto_transformer}, or the GPU resource management perspective~\cite{gslice,mig_serving,stanf}. However, achieving the ult-most efficiency for multi-tenant DL computing is more challenging as it needs to thoroughly consider the differences between single vs. multi-tenant DL inference, and requires multi-layer DL optimization or full-stack co-optimization.
As so, there is a great need for a systematical review of opportunities and challenges on multi-tenant DL inference optimization.

\textit{Our survey is the first work} that thoroughly analyzes the multi-tenant GPU scheduling problem, summarizes the major differences in single- vs. multi-tenant computing optimization, and reveals the emerging opportunities and potential benefits of multi-tenant DL inference on GPUs. To ease the understanding, our work also draws some experience from pevious DL computing stacks, and compares single vs multi-tenant DL inference on GPU from a hierarchical perspective. 





\vspace{0.5mm}
\textbf{Single vs Multi-Tenant: A Hierarchical Comparison.} Traditional single-tenant DL compilers (Figure~\ref{fig:1} (c)) already include multi-layer optimization: algorithm-level compression~\cite{feature_prune, filter_prune, connect_prune}, graph-level rewriting~\cite{taso,ansor,horizon}, runtime scheduling~\cite{ios,nimble} and kernel tuning~\cite{tvm,winograd,direct_conv,trt}, etc. However, optimizations targeting at single-tenant are usually ill-fitted for multi-tenant inference with following examples.

From the top graph level, multi-tenant DL inference workloads represented in directed acyclic graphs (DAGs) come with significantly higher volume of multi-model operators than single model. This incurs many non-mergeable operators and exposes much larger scheduling space.
Another example in the lower kernel level is TVM~\cite{tvm}. As one of the most competitive kernel auto-tuning framework, TVM comes with a built-in assumption of single-tenant execution setting, the tuning configuration of which aims to saturate all SMs and memory bandwidth of the GPU.
Such assumption and the single-tenant targeted tuning however becomes unsuitable for multi-kernel concurrent execution with partial resource available for each kernel. According to~\cite{mt_tvm}, the maximum throughput gap comparing single-tenant vs. multi-tenant tuned configurations for the same computing kernel could reach $5\times$ difference.





Further down to the GPU hardware, multi-tenant DL inference requires many scheduling support to address problems like inter-tenant interference, such as dedicated GPU hardware primitives for resource partitioning, isolation and allocation, etc.
With the current trends towards multi-tenant applications, GPU vendors like Nvidia have recently released many new features including Multi-Stream~\cite{stream}, Multi-Process Service (MPS)~\cite{mps}, Multi-Instance GPU (MIG)~\cite{mig} and virtual GPUs (vCS)~\cite{vcs} to support both runtime scheduling and resource management, which exposes new research opportunities for multi-tenant DL scheduling and optimization.

With such differences considered, this work will adopt such a view of optimization stack to thoroughly analyze the multi-tenant challenges and introduce emerging works.

\vspace{3mm}
By reviewing the emerging challenges, opportunities, and research works on multi-tenant DL computing on the GPU, we hope this survey could motivate more design and innovations in this promising new domain. 
	The remaining paper is organized as follows:
	Section~\ref{sec:bg} introduces the novel challenges and opportunities in multi-tenant GPU computing stack and a high-level overview of current vendor GPU support.
	Section~\ref{sec:misd} summarizes recent research works for multi-tenant computing in detail.
	We then give our vision and insights in Section~\ref{sec:vision}.
	Section~\ref{sec:con} concludes this paper.


%% file: _txt/2_def.tex
\section{\textbf{Challenges \& Opportunities\\ for Multi-Tenant Computing on GPU}}
\label{sec:bg}

In this section, we first characterize the major differences between single- vs. multi-tenant DL computing optimization through the full DL computing stack. We then introduce the recently-released GPU features, such as Stream, MPS, MIG, which provide important fundamental backend support for multi-tenant computing optimization. 

\subsection{\textbf{Challenges for Multi-Tenant DL Computing}}

\begin{figure}[!tb]
  \centering
  \vspace{-1mm}
  \includegraphics[width=3.4in]{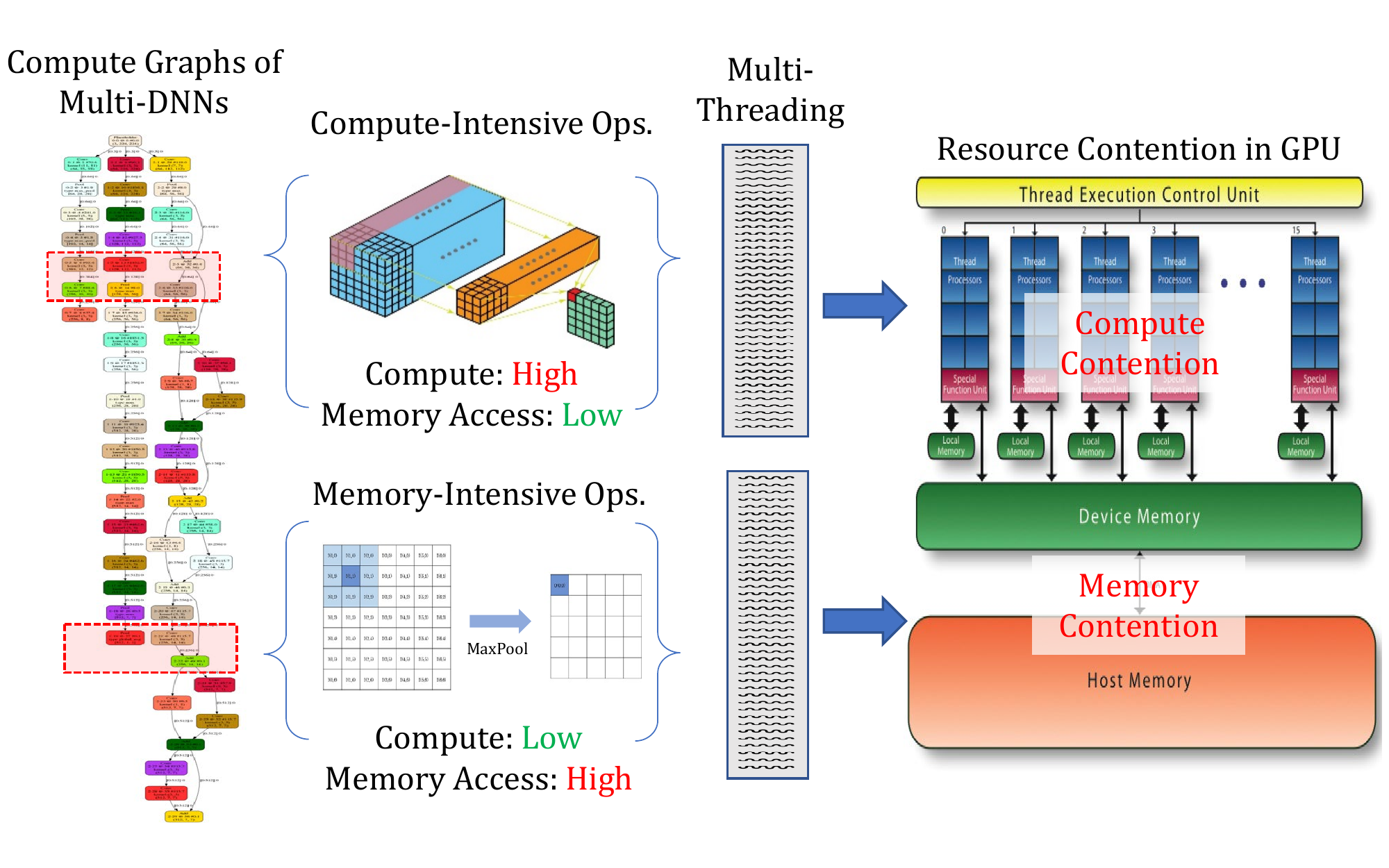}
  \vspace{-5mm}
  \caption{Multi-Tenant DL Computing Challenges.}
  \label{fig:challenges}
  \vspace{-4mm}
\end{figure}

Traditional DL computing optimization in full stack often expands in \circled{1} \textit{service-level} orchestration~\cite{dlis}, \circled{2} \textit{graph-level} optimization~\cite{taso, mlsys}, \circled{3} \textit{runtime-level} scheduling~\cite{ios}, \circled{4} \textit{kernel-level} tuning~\cite{tvm, xla} and \circled{5} \textit{resource-level} management~\cite{gslice,stanf}. 
Although there are many previous works for computing optimization in these difference levels, multi-tenant computing shows dramatic characteristics that make these methods ill-fitted. According the the same optimization stack, we summarize the major differences in Table~\ref{table:compare} and analyze the computing challenges in Figure~\ref{fig:challenges}.

\begin{table}[!b]
\small
\vspace{-6mm}
\caption{Challenges for Multi-Tenant Optimization.}
\vspace{-1mm}
\setlength{\tabcolsep}{1mm}{
\begin{adjustbox}{width=0.48\textwidth}
\renewcommand\arraystretch{1.4}
\begin{tabular}{l|l|l|l}
\hline \hline
\multicolumn{2}{c|}{\textbf{Full Optimization Stack}} & \textbf{Single-Tenant}                    & \textbf{Multi-Tenant}                           \\ \hline \hline
    & Co-location                        & {\color[HTML]{32CB00} \textbf{No}}      & {\color[HTML]{FE0000} \textbf{Yes}} \\ \cline{2-4} 
\multirow{-2}{*}{\circled{1} \color[HTML]{8A2BE2} \textbf{Service-level}}   & Interference       & {\color[HTML]{32CB00} \textbf{No}}         & {\color[HTML]{FE0000} \textbf{High Interference}}               \\ \hline
{\circled{2} \color[HTML]{98817B} \textbf{Graph-level}} & DAG(s)               & {\color[HTML]{32CB00} \textbf{Mostly Seq.}}      & {\color[HTML]{FE0000} \textbf{Seq. + Parallel}} \\ \hline
                   & Parallelism                     & {\color[HTML]{32CB00} \textbf{Limited}}   & {\color[HTML]{FE0000} \textbf{Extensive}}       \\ \cline{2-4} 
\multirow{-2}{*}{\circled{3} \color[HTML]{FFBF00} \textbf{Runtime-level}} & Complexity   & {\color[HTML]{32CB00} \textbf{Low}}        & {\color[HTML]{FE0000} \textbf{High}}              \\ \hline
                   & Resource Usage                  & {\color[HTML]{32CB00} \textbf{Exclusive}} & {\color[HTML]{FE0000} \textbf{Shared}}          \\ \cline{2-4} 
\multirow{-2}{*}{\circled{4} \color[HTML]{6495ED} \textbf{Kernel-level}}  & Tuning Objective & {\color[HTML]{32CB00} \textbf{100\% util.}} & {\color[HTML]{FE0000} \textbf{$x$\% partial util.}} \\ \hline
\multicolumn{1}{c|}{\circled{5} \color[HTML]{03C03C} \textbf{Resource-level}} &
  Management &
  {\color[HTML]{32CB00} \textbf{No}} &
  {\color[HTML]{FE0000} \textbf{Resource Partition}} \\ \hline \hline
\end{tabular}
\end{adjustbox}}
\label{table:compare}
\end{table}
\normalsize

\vspace{1mm}
\noindent \circled{1} \textbf{Service-level:} AI-centric cloud services handle millions of service queries simultaneously~\cite{fb_center}. With the massive computing capacity of GPUs, multiple DL queries could be \textit{strategically co-located for efficient concurrent execution}, which is one key difference between multi-tenant GPU computing versus traditional CPU multi-tasking. 
By allowing the resource sharing among concurrent DL workloads, the service providers could potentially improve the GPU resource utilization and reduce cost of ownership (COO) like infrastructure and power cost especially for large-scale data centers~\cite{mig}.

However, the challenges remains for strategic co-location like that the \textit{inter-tenant interference}~\cite{stanf} could happen and degrade the quality of service such as service-level objectives (SLA) of tail latency and throughput.
This could become worse with increased number of co-located workloads and degrade the overall serving throughput.
Therefore, there are many recent \textit{service-level orchestration} works~\cite{stanf,irina,prema} that design different heuristic-based, modeling-based or prediction-based mechanisms to conduct strategic co-location for efficient multi-tenant computing on GPUs.

\vspace{1mm}
\noindent \circled{2} \textbf{Graph-level:} DNNs with many operators are commonly represented as directed acyclic graphs (DAGs), which use nodes to represent operators and edges to represent the data flow and dependency~\cite{taso}. 
Single-model DAGs are usually \textit{sequential with limited parallelism} like VGG, ResNets, MobileNets and EfficietNets, which have only one or two branches and thus exposes small scheduling space~\cite{advanced}. 

By contrast, multi-tenant DL computing with multiple parallel DAGs usually have \textit{extensive inter-operator parallelism}, as shown in Figure~\ref{fig:challenges}, which enables more flexible inter-operator scheduling among tenants~\cite{iccad}. Certain challenges exist in such graph scheduling such as the increased complexity in larger number of operators and scheduling space, and the more complex GPU resource contention analysis, etc.

\vspace{1mm}
\noindent \circled{3} \textbf{Runtime-level:} Previously due to the limited inter-operator parallelism, only a few works~\cite{ios, auto_transformer} touch upon runtime-level scheduling such as certain multi-branch models like Inception, NasNets and Transformers. 
These works leverage certain GPU runtime primitives (\eg, Nvidia multi-stream~\cite{stream}) for concurrent operator scheduling, many of which however incur large runtime overheads. For example, multi-stream synchronization forces all streams to wait/stall until the last stream finishes its workloads~\cite{stream}.
Multi-tenant scheduling tends to suffer more from such overheads with the increased number of operators and scheduling complexity.
Due to the increased attention in GPU multi-tenant scheduling, GPU vendors have recently released a series of important features such as CUDA graphs~\cite{graph} to address such scheduling overheads. 

\vspace{1mm}
\noindent \circled{4} \textbf{Kernel-level:} Kernel configurations such as loop tiling, thread blocking, memory colasing, \etc could significantly influence each operator's computing efficiency. Previous single-tenant kernel-level works like TVM~\cite{tvm} and TF-XLA~\cite{xla} try to find the best configuration that can saturate the GPU resource, i.e., \textit{exclusive resource usage}. 
However, as multi-tenant DNNs \textit{share the underlying resource}, kernels optimized for single-tenant settings can easily become sub-optimal for multi-tenant scenarios. Recently, there are certain works that show multi-tenant DL computing should optimize kernel configurations according to its available resource ratio during practical execution~\cite{mt_tvm}, which shows a 5$\times$ throughput difference.

\vspace{1mm}
\noindent \circled{5} \textbf{Resource-level:}
To achieve \textit{adaptive multi-tenant resource partitioning and provisioning}, it asks for both strategic design and hardware support. 
The first challenge for such adaptive resource provisioning lies in the \textit{DL workload dynamics}~\cite{iccad}. Multiple DL models with different deep structures have highly non-stable computing/memory requirements, making the inter-model resource sharing and competition highly dynamic and thus hard to determine the optimal resource partitioning. 
On the other hand, adaptive resource management requires the \textit{flexible GPU reconfiguration capability}.
Although there are certain adaptive resource provisioning features (\eg, Nvidia multi-process service, multi-instance GPU~\cite{mps, mig}) that supports resource partitioning, the reconfiguration process requires non-negligible time (\eg, several $ms$), which is a major limitation of the recent resource scheduling works~\cite{gslice, stanf}.

\begin{figure}[!tb]
  \centering
  \vspace{5mm}
  \includegraphics[width=3.4in]{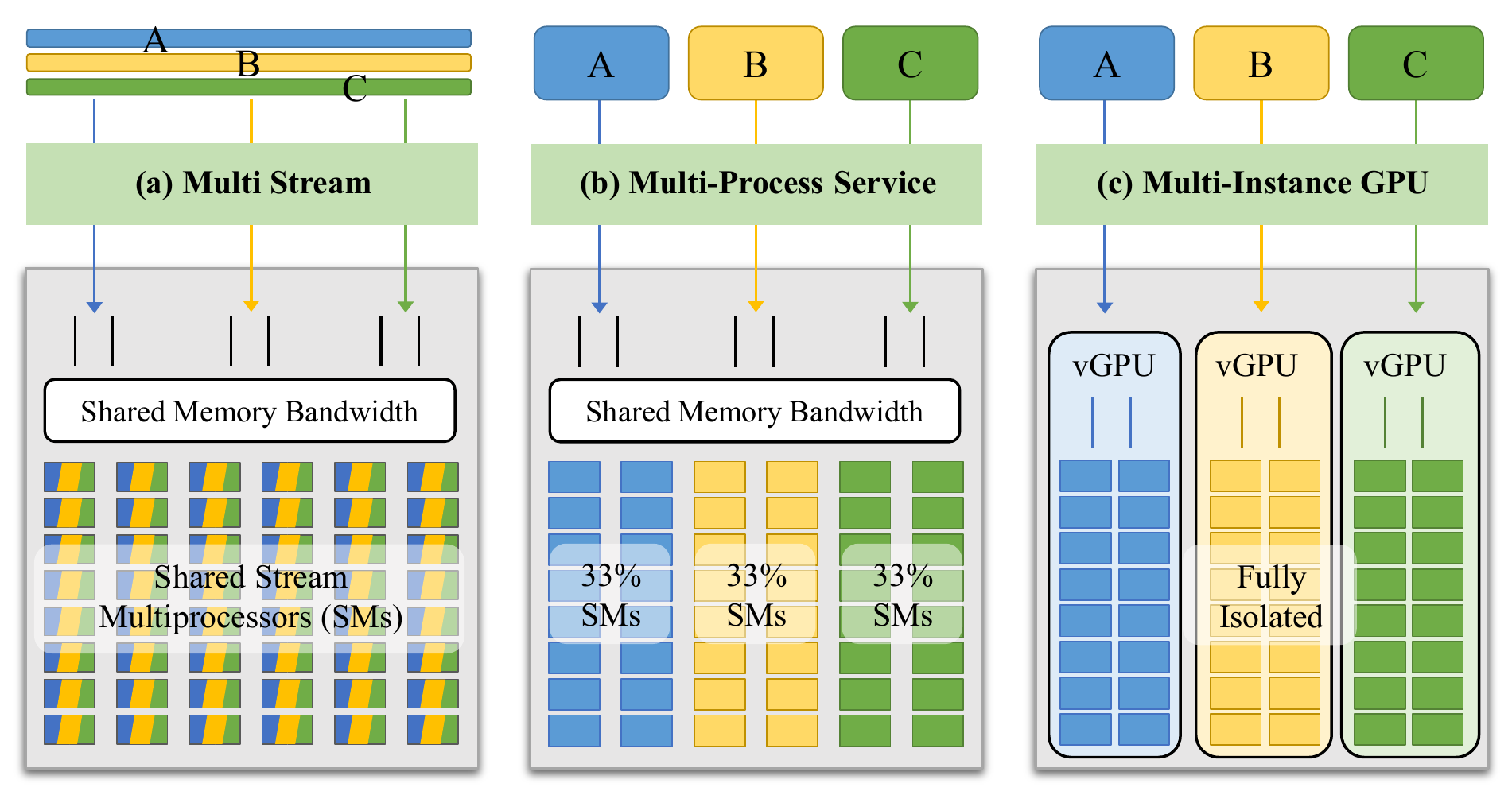}
  \vspace{-2mm}
  \caption{Multi-Stream, MPS and MIG Illustration.}
  \label{fig:mst_mps_mig}
  \vspace{-3.9mm}
\end{figure}

\subsection{\textbf{Emerging Multi-Tenant Computing Opportunities}}



Previously, one important reason that hinders the adoption and development of multi-tenant DL computing on GPU is the insufficient hardware scheduling mechanisms. 
But with the increasing attention in this topic, GPU vendors like Nvidia release certain new GPU multi-tenant features to support multi-tenant DL inference, which provides \textit{great opportunities for multi-tenant scheduling}.
The multi-tenant GPU scheduling features could be categorized into two major types: software-level and hardware-level support.

\vspace{1mm}
\noindent \circled{1} \textbf{Software Approach:}
The very first GPU multi-tasking feature is the \textbf{\textit{Multi-Stream}} mechanism~\cite{stream} supported in the Fermi GPU architecture (Figure~\ref{fig:mst_mps_mig} (a)).
  As a software-based programming model, a stream can contain a sequence of issued operations that execute on the GPU.
  Operations in different streams could run concurrently and share the underlying GPU resources like SMs~\cite{stream}.
  The similar feature \textbf{\textit{Hyper-Q}}~\cite{hyperq} is introduced in the Kepler GPU architecture (2013) that expands previous 16-way to 32-way hardware kernel queues for higher concurrency support.
  Along with the concurrency support, the CUDA library also releases certain scheduling APIs like DeviceSync, StreamWait, etc. to support more fine-grained scheduling capability~\cite{cuda}.
  These software-level APIs provide valuable multi-tenant GPU scheduling mechanisms, based on which many recent works have started to explore the fine-grained DL operator-level scheduling techniques~\cite{iccad,ios}.

\begin{table}[!tb]
\footnotesize
\caption{GPU Support for Multi-Tenant Computing.}
\vspace{-1mm}
\begin{adjustbox}{width=0.48\textwidth}
\renewcommand\arraystretch{1.4}
\setlength{\tabcolsep}{1.5mm}{
\begin{tabular}{lccc}
\hline \hline
\textbf{Mechanisms}   & \textbf{Stream} & \textbf{MPS}   & \textbf{MIG} \\ \hline \hline
\textbf{Partition Type}        & No              & Logical        & Physical     \\ \hline
\textbf{Max Partition}         & Unlimited       & 48             & 7            \\ \hline
\textbf{SM Isolation}          & No              & By Percentage  & Yes          \\ \hline
\textbf{Mem BW Isolation}      & No              & No             & Yes          \\ \hline
\textbf{Performance QoS}       & No              & Partial        & Yes          \\ \hline
\textbf{Reconfiguration}       & Dynamic         & Process Launch & When Idle    \\ \hline \hline
\end{tabular}} 
\end{adjustbox}
\vspace{1mm}
\label{tab:mst_mps_mig}
\end{table}

\begin{table*}[!t]
\scriptsize
\caption{Recent Works on Multi-Tenant Computing Optimization (JCT: job completion time, SLA: service-level agreement).}
\vspace{-1mm}
\renewcommand\arraystretch{1.2}
\setlength{\tabcolsep}{3.7mm}{
\begin{tabular}{lllll}
\hline
\multicolumn{1}{c}{\textbf{Ref.}} &
  \multicolumn{1}{c}{\textbf{Hardware}} &
  \multicolumn{1}{c}{\textbf{Perspective}} &
  \multicolumn{1}{c}{\textbf{Algorithm/Strategy}} &
  \multicolumn{1}{c}{\textbf{Improvement/Achievement}} \\ \hline
Inter-Aware~\cite{stanf} &
  GPU &
  DL Service-level Orchestration &
  \begin{tabular}[c]{@{}l@{}}$\bullet$ ML-based Interference Predictor \\ $\bullet$ Proactive Query Scheduler \end{tabular} &
  \begin{tabular}[c]{@{}l@{}}$\bullet$ Reducing Job Interference \\$\bullet$ Enhancing Serving Throughout\end{tabular} \\ \hline
Irina~\cite{irina} &
  GPU &
  DL Service-level Orchestration &
  \begin{tabular}[c]{@{}l@{}}$\bullet$ Online Query Scheduler\\$\bullet$ Heuristic-based Preemption\\$\bullet$ Concurrent Execution \& Batching\end{tabular} &
  $\bullet$ Reducing Client-Side JCT \\ \hline
PREMA~\cite{prema} &
  NPU &
  DL Service-level Orchestration &
  \begin{tabular}[c]{@{}l@{}}$\bullet$ Online Query Scheduler\\$\bullet$ Heuristic-based Preemption\end{tabular} &
  \begin{tabular}[c]{@{}l@{}}$\bullet$ Reduced High-Priority Job JCT\\$\bullet$ Maintaining Low-Priority SLA\end{tabular} \\ \hline
Runtime-Aware~\cite{iccad} &
  GPU &
  Graph \& Runtime-level Scheduling &
  \begin{tabular}[c]{@{}l@{}}$\bullet$ Multi-Model DAG Rewriting \\$\bullet$ ML-based Scheduling Search \\$\bullet$ Multi-Stream Runtime Scheduling \end{tabular} &
  $\bullet$ Reduced Inference Latency \\ \hline
Spatial-Tune~\cite{mt_tvm} &
  GPU &
  Kernel-level Auto-Tuning &
  \begin{tabular}[c]{@{}l@{}}$\bullet$ MPS-based Resource Allocation\\$\bullet$ Partial-Resource Kernel Tuning\end{tabular} &
  \begin{tabular}[c]{@{}l@{}}$\bullet$ Enhanced Kernel Performance\\$\bullet$ Reduced Inter-kernel Interference\end{tabular} \\ \hline
GSlice~\cite{gslice} &
  GPU &
  Resource-level Management &
  \begin{tabular}[c]{@{}l@{}}$\bullet$ MPS-based Resource Partitioning\\$\bullet$ Adaptive Batching\end{tabular} &
  \begin{tabular}[c]{@{}l@{}}$\bullet$ Enhanced Serving Throughput\\$\bullet$ Maintaining SLA\end{tabular} \\ \hline
Spatial-Partition~\cite{kaist} &
  GPU &
  Resource-level Management &
  \begin{tabular}[c]{@{}l@{}}$\bullet$ MPS-based Resource Partitioning\\$\bullet$ Interference-aware Scheduling\end{tabular} &
  \begin{tabular}[c]{@{}l@{}}$\bullet$ Enhanced Serving Throughput\\$\bullet$ Maintaining SLA\end{tabular} \\ \hline
MIG-Serving~\cite{mig_serving} &
  GPU &
  Resource-level Management &
  \begin{tabular}[c]{@{}l@{}}$\bullet$ MIG-based Resource Reconfiguration\\$\bullet$ Fast \& Slow Query Scheduling\end{tabular} &
  \begin{tabular}[c]{@{}l@{}}$\bullet$ Enhanced Serving Throughput\\$\bullet$ Maintaining SLA\end{tabular} \\ \hline
Planaria~\cite{plana} &
  \begin{tabular}[c]{@{}l@{}}Systolic\\ Arrays\end{tabular} &
  Resource-level Management &
  $\bullet$ Architecture Reconfiguration &
  \begin{tabular}[c]{@{}l@{}}$\bullet$ Enhanced Serving Throughput\\$\bullet$ Reduced Energy Consumption\end{tabular} \\ \hline
\end{tabular}}
\label{table:summary}
\vspace{-3mm}
\end{table*}
\normalsize

\vspace{1mm}
\noindent \circled{2} \textbf{Hardware Approach:}
Besides the software support, NVIDIA recently also releases advanced hardware-level resource management mechanisms to support \textit{flexible resource allocation, isolation and virtualization}.
These resource management methods can be categorized into two types: \textit{logical} and \textit{physical}.
\textbf{\textit{Multi-Process Service (MPS)}}~\cite{mps} is a logical resource partitioning mechanism (Figure~\ref{fig:mst_mps_mig} (b)) that allows user to partition the streaming multi-processors (SMs) and allocate them to different processes, for example, 30\%, 70\% to two concurrent processes. 
  Such partition is done by the software-based process-to-SM mapping scheduling and thus considered logical.
  Notably, although MPS enables logical SM partitioning, other GPU resources like memory bandwidth are not partitioned and thus MPS cannot fully avoid the inter-process resource competition and interference. 
  To address this, the recently introduced \textbf{\textit{Multi-Instance GPU (MIG)}}~\cite{mig} on Ampere architecture enables physical partitioning of both SMs and memory bandwidths through dedicated GPU architecture design (Figure~\ref{fig:mst_mps_mig} (c)).
  Such physical partition ensures fully isolated resources, and thus no interference can happen between different processes.
  MIG support splitting one A100 GPU into seven fully isolated GPU instances.
  Meanwhile, it provides certain \textit{reconfiguration capability} when the GPU is fully or partial idle. For example, one A100 could be split into three instances with the ratio of 4:2:1 and then reconfigured to be 3:3:1, etc~\cite{mig}. 
  More detailed comparison of Stream, MPS, MIG could be found in Table~\ref{tab:mst_mps_mig}.

%% file: _txt/3_tech.tex
\section{\textbf{Multi-Tenant Computing Optimization: \\Design and Innovations}}
\label{sec:misd}

Based on the former challenges and opportunities, we review the emerging works tackling the multi-tenant scheduling optimization from different perspectives. We summarize these works in Table~\ref{table:summary}. From a top-down view, these works are categorized into several levels, \ie, from \textit{DL service-level orchestration}, \textit{graph \& runtime-level scheduling}, to \textit{kernel-level auto-tuing} and then \textit{GPU resource-level management}. 


\subsection{\textbf{Service-level Orchestration}}

DL service-level orchestration is an important feature in large-scale data centers to improve the GPU utilization. 
As the top-most scheduling level, such orchestration usually regards one service query as the basic scheduling unit. This reduces the scheduling complexity as there is no need to consider the intra-DNN model structure details (operators and graphs). 
One example is the Microsoft Deep Learning Inference Service (DLIS) system~\cite{dlis}. 
The service orchestrator characterizes different models’ resource requirements and then strategically places one or multiple queries onto hosts through the service router. 
Therefore, it could maximize the served queries per second (QPS) while ensure little inter-query interference so as to maintain similar tail latency.

However, designing a proper co-location strategy or system is a non-trivial task.
For example, one challenging factor is the {serving dynamics}, \ie, undetermined arrival rates and/or distribution of incoming DL queries, different RNN/LSTMs queries with varied inputs and control states.
  Distinct from static workloads that we can get the full information, such dynamic scenarios require us to either utilize historical data or predict the future workload dynamics. 
  PREMA~\cite{prema} proposed a predictive multi-task DNN scheduling algorithm that combines off-line records and online token-based job scheduling to determine the best multi-tenant co-location strategy. 
  

Another challenge in multi-tenant co-location is how to accurately predict the inter-model resource interference. This is a critical factor in ensuring QoS such as tail latency. \cite{stanf} trained a ML-based latency degradation predictor under co-location using offline-profiled hardware-level features such as SM and DRAM usage, PCIE read/write BW, buffer usage, etc. Then the latency degradation predictor is used to evaluate the model placement's potential influence for each query. 

However, these works have certain scalability issues as they mostly targeted at static model types, hardware types, etc., which may not be suitable for dynamic workloads.
Meanwhile, as each DNN can have many operators (\eg, layers) that have fluctuated resource consumption, such coarse-grained scheduling (with one entire query as the basic unit) may suffer from resource under-utilization/contention occasionally and thus still hinders the QoS.

%


\subsection{\textbf{Graph and Runtime-level Scheduling}}

Graph and runtime-level scheduling could help address one of the aforementioned challenge of coarse-grained granularity by enabling more fine-grained scheduling, \eg, the DNN operators. 
  This could be done by leveraging the GPU software-level support such as the multi-stream mechanism and scheduling APIs.
For example, \cite{iccad} propose an ML-based scheduling strategy for multi-tenant DNN exeuction acceleration.
It first abstracts multiple DNN's computation graph with all operators into a global intermediate representation (IR), which enables flexible resource sharing between different tenants so as to improve the utilization. 
To find the optimal concurrent operator execution strategy in the huge scheduling space, they design a ML-based auto-search method by defining three main factors: scheduling search space, profiling-guided latency cost model, and the ML search algorithm.
Based on offline profiling records, the search algorithm could find the best scheduling for optimal GPU utilization and throughput. 

Such graph and runtime-level operator scheduling could usually achieve better performance due to the fine-grained design, but they also face more scalability issues, \eg, when the number of co-located workloads increase to very large. Meanwhile, it also applies to static or known multi-tenant workload only, which cannot address dynamic model types.

\subsection{\textbf{Resource-Level Management}}

Besides the aforementioned works, another optimization perspective to solve the inter-tenant inference is to conduct fine-grained resource managing~\cite{gslice, plana}. For example, spatial partitioning and allocation of GPU resources to different DL workloads could isolate different jobs' resource (\eg, stream multiprocessors (SMs), memory bandwidths), thus avoiding the job interference in the hardware resource level.
However, as we introduced before, achieving fine-grained resource partitioning is non-achievable until recently GPU vendors release a series of resource sharing and partitioning support like multi-streams, multi-process services (MPS~\cite{mps}) and multi-instance GPU (MIG~\cite{mig}).
Most recent resource-level management works are built upon these technologies. 

For example, GSlice~\cite{gslice} uses MPS to conduct adaptive SM partitioning for different DNN jobs. They design a self-learning method to dynamically adjust the GPU resource allocation ratio for each workload and thus avoid interference among co-located DL workloads and maximize the throughput.
\cite{kaist} utilizes similar spatial partitioning mechanism by MPS while additionally combines temporal scheduling strategies. 
MIG-Serving~\cite{mig_serving} is the most recent work that adopts the newly-released MIG feature on A100 to achieve spatial resource management for multi-tenant scheduling. 

However, such spatial resource partitioning solutions also have a intrinsic limitation that is the \textit{inflexible re-configuration} when the workloads change and requires resource partitioning adjustment. For GPUs, re-configuring the resource partitioning requires certain amount of time (\eg, tens of $ms$ or more), which can be even larger than one DL inference workloads' processing time. Therefore, re-configuring frequently is not practical and thus limits such solutions' performance when facing dynamic workloads. \cite{gslice} tries to reduce the stall caused by reconfiguration time of MPS by utilizing a standby/shadow process. However, the minimum time for switching one partitioning configuration to another one still cost several seconds, which is still non-negligible in online serving.

\subsection{\textbf{Potential Directions for Remaining Challenges}}
\label{sec:misd_future}

\noindent \circled{1} \textbf{ML-based Prediction and Online Learning:}
To address the problem of service dynamics, using ML-based predictive model (\eg, reinforcement learning, LSTM, \etc) is one promising direction, which can potentially predict the future queries trend and guide the overall scheduling. The ML-based model can be initially trained offline by historical serving records. During the online serving process, active learning and continual learning~\cite{active,lifelong} using the latency/throughput as feedback can be potentially utilized to improve the predictive accuracy and the scheduling effectiveness consistently.


Another way of leveraging ML-based prediction is to conduct light-weight modeling to predict the latency degradation under different multi-model and hardware combinations so that the scheduler can make better decision regarding the latency SLA constraints. For example, the work~\cite{stanf} built a ML model to predict the latency of multi-model inference cases on different machines. As the effectiveness of the final scheduling solution highly depends on the modeling accuracy, the scalability and generality issue across hardware/model types needs to be addressed, which can be also very challenging.

\begin{figure*}[!t]
  \centering
  \includegraphics[width=5in]{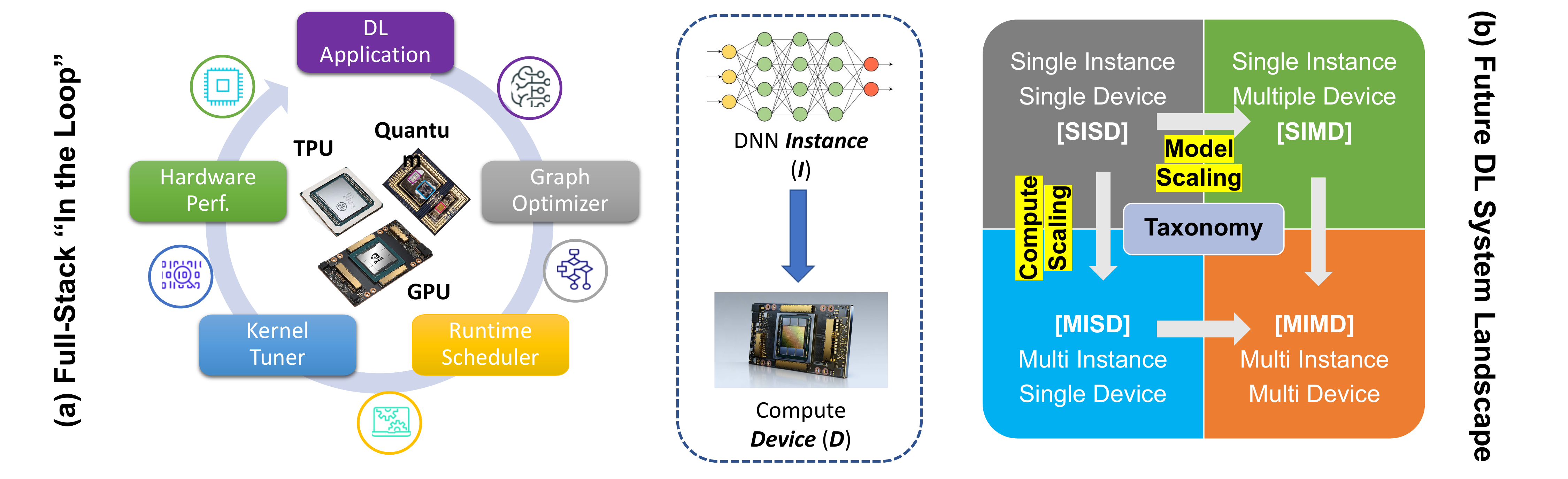}
  \vspace{-1mm}
  \caption{The future trends to a larger scale DL system.}
  \vspace{-3mm}
  \label{fig:taxonomy}
\end{figure*}


\vspace{1mm}
\noindent \circled{2} \textbf{Software-Hardware Co-Scheduling:}
The software and hardware scheduling could be complementary to provide both high job scheduling flexibility and strict resource isolation. 
There are some recent works that adopt such a temporal-spatial combined perspective. 
\cite{kaist} uses MPS to conduct resource partitioning and then implements a heuristic-based task scheduler to find the appropriate mapping between the DNN queries and gpu partitions. 
In addition to that, software-hardware scheduling could also be leveraged to alleviate certain re-configuration overhead. 
For example, it's potential to conduct software-based scheduling within a partitioned GPU slice, \eg, combining multi-stream with MIG. In this way, fine-grained scheduling could be achieved without re-partitioning the entire GPU, avoiding the reconfiguration overhead. 
%

%% file: _txt/4_future.tex









\section{\textbf{Towards Large-Scale DL Computing:\\ Vision and Insights}}
\label{sec:vision}


\subsection{\textbf{Architecture Design with ``Full Stack in the Loop"}}

The fast development of multi-tenant DL computing brings many challenges for the system stack optimizations.  Besides for the GPU only, this also enlightens the other DL-oriented hardware architecture designers (\eg, TPU, chiplet, neuromorphic and quantum-based accelerators) to optimize for \textit{flexibility and agility} facing a rapidly changing DL application landscape.
Specifically, one important future trend is the \textit{``full stack in the loop"}, i.e., to remove the boundaries in the vertical DL system stack and conduct full-stack integration to strive for both optimal performance and flexibility. 

One example in this line of efforts is the DL compiler renovation by tvm unity~\cite{tvm_talk}, as shown in Figure~\ref{fig:taxonomy} (a). Current DL computing stack conducts separate layer-wise optimization (graph-runtime-kernel-resource) and single directional \textit{top-down} deployment. This however prohibits feedback loops and cross-layer interactions for SW/HW co-compiling based on model workloads and hardware characteristics. Unifying the abstraction between layers would thus greatly facilitate the new full-stack optimization as a loop, not only for multi-tenant computing, but also for future wider DL application.

Another example is the increasing attention in the versatile and flexible chiplet-based SW/HW co-design~\cite{chiplet2, chiplet1} that uses multi-chip-modules (MCMs). Compared to traditional large monolithic die, such MCM combines smaller chiplets into a larger system and substantially reduces fabrication and design costs. However, it requires thorough application awareness to optimize the chip design and overall performance. As so, such chiplet modules could also greatly benefit from the full-stack-in-the-loop architecture of DL computing.

\vspace{1mm}
\subsection{\textbf{The Future Large-Scale DL System Landscape}} 
Multi-tenant DL computing is a natural generalization result due to the significant \textit{computing scaling} trend of GPU.
However, recently another \textit{model scaling} trend is observed, that is, designing and training super-scale AI models for general intelligence. For example, the recent SOTA giant AI model Megatron-NLG~\cite{mt_nlg} has reached 530 billions of parameters and requires tens of GPUs to conduct multi-node distributed inference.  If we take such \textit{model scaling} into consideration, even more new DL computing modes could be observed and enrich the future DL \& system landscape.

We describe the future large-scale DL system landscape by using a taxonomy in Figure~\ref{fig:taxonomy} (b). 
Using {Instance (I)} to denote one DNN model and {Device (D)} to denote the hardware, traditional DL system mostly comes within the \textit{Single Instance Single Device (SISD)} domain and only constitute the top-left quarter in the full spectrum.
Multi-tenant computing emerges as the \textit{Multiple Instances Single Device (MISD)} with the computing scaling trend, as we summarized in this servery. 

Whereas diagonally, with the model scaling trend, the \textit{Single Instance Multiple Devices (SIMD)} interaction mode also emerges and is attracting more attention such as distributed inference for super-scale giant models~\cite{fb,deepthing,hotedge} including language, recommendation models, etc.
Finally, \textit{Multiple Instances Multiple Devices (MIMD)} computing would eventually combine all these modes and become a practical needs for future DL-centric data center optimization.

%% file: _txt/6_conclusion.tex
\section{\textbf{Conclusion}}
\label{sec:con}

Deep Learning-based intelligence creates a wide-spectrum of applications featured with substantial complexity such as multi-modality and multi-tasking. GPU is one major type of DL accelerators that democratized such complex DL applications in both cloud and edge. Meanwhile, its gen-by-gen capacity demonstrates exponential scaling. With such application complexity and GPU capacity scaling,  multi-tenant DL computing is emerging as an effective computing paradigm on GPU to enhance the throughput, resource utilization, and power efficiency.
This survey summarizes and categorizes the emerging challenges and optimization opportunities for multi-tenant DL inference on GPU following a hierarchical comparison with traditional single-tenant optimization. 
  We hope that this survey could shed lights on new perspectives and novel works in future large-scale DL system optimization.


%% file: survey.bbl

\begin{thebibliography}{49}


\ifx \showCODEN    \undefined \def \showCODEN     #1{\unskip}     \fi
\ifx \showDOI      \undefined \def \showDOI       #1{#1}\fi
\ifx \showISBNx    \undefined \def \showISBNx     #1{\unskip}     \fi
\ifx \showISBNxiii \undefined \def \showISBNxiii  #1{\unskip}     \fi
\ifx \showISSN     \undefined \def \showISSN      #1{\unskip}     \fi
\ifx \showLCCN     \undefined \def \showLCCN      #1{\unskip}     \fi
\ifx \shownote     \undefined \def \shownote      #1{#1}          \fi
\ifx \showarticletitle \undefined \def \showarticletitle #1{#1}   \fi
\ifx \showURL      \undefined \def \showURL       {\relax}        \fi
\providecommand\bibfield[2]{#2}
\providecommand\bibinfo[2]{#2}
\providecommand\natexlab[1]{#1}
\providecommand\showeprint[2][]{arXiv:#2}

\bibitem[\protect\citeauthoryear{Bai, Yao, Sun, and Yu}{Bai
  et~al\mbox{.}}{2021}]%
        {auto_transformer}
\bibfield{author}{\bibinfo{person}{Yang Bai}, \bibinfo{person}{Xufeng Yao},
  \bibinfo{person}{Qi Sun}, {and} \bibinfo{person}{Bei Yu}.}
  \bibinfo{year}{2021}\natexlab{}.
\newblock \showarticletitle{AutoGTCO: Graph and Tensor Co-Optimize for Image
  Recognition with Transformers on GPU}. In \bibinfo{booktitle}{\emph{2021
  IEEE/ACM International Conference On Computer Aided Design (ICCAD)}}.
\newblock


\bibitem[\protect\citeauthoryear{Chen, Moreau, Jiang, Zheng, Yan, Shen, Cowan,
  Wang, Hu, Ceze, et~al\mbox{.}}{Chen et~al\mbox{.}}{2018}]%
        {tvm}
\bibfield{author}{\bibinfo{person}{Tianqi Chen}, \bibinfo{person}{Thierry
  Moreau}, \bibinfo{person}{Ziheng Jiang}, \bibinfo{person}{Lianmin Zheng},
  \bibinfo{person}{Eddie Yan}, \bibinfo{person}{Haichen Shen},
  \bibinfo{person}{Meghan Cowan}, \bibinfo{person}{Leyuan Wang},
  \bibinfo{person}{Yuwei Hu}, \bibinfo{person}{Luis Ceze}, {et~al\mbox{.}}}
  \bibinfo{year}{2018}\natexlab{}.
\newblock \showarticletitle{$\{$TVM$\}$: An automated end-to-end optimizing
  compiler for deep learning}. In \bibinfo{booktitle}{\emph{13th $\{$USENIX$\}$
  Symposium on Operating Systems Design and Implementation ($\{$OSDI$\}$ 18)}}.
  \bibinfo{pages}{578--594}.
\newblock


\bibitem[\protect\citeauthoryear{Choi, Lee, Kim, Park, Kwon, and Huh}{Choi
  et~al\mbox{.}}{2021}]%
        {kaist}
\bibfield{author}{\bibinfo{person}{Seungbeom Choi}, \bibinfo{person}{Sunho
  Lee}, \bibinfo{person}{Yeonjae Kim}, \bibinfo{person}{Jongse Park},
  \bibinfo{person}{Youngjin Kwon}, {and} \bibinfo{person}{Jaehyuk Huh}.}
  \bibinfo{year}{2021}\natexlab{}.
\newblock \showarticletitle{Multi-model Machine Learning Inference Serving with
  GPU Spatial Partitioning}.
\newblock \bibinfo{journal}{\emph{arXiv preprint arXiv:2109.01611}}
  (\bibinfo{year}{2021}).
\newblock


\bibitem[\protect\citeauthoryear{Choi and Rhu}{Choi and Rhu}{2020}]%
        {prema}
\bibfield{author}{\bibinfo{person}{Yujeong Choi} {and} \bibinfo{person}{Minsoo
  Rhu}.} \bibinfo{year}{2020}\natexlab{}.
\newblock \showarticletitle{Prema: A predictive multi-task scheduling algorithm
  for preemptible neural processing units}. In \bibinfo{booktitle}{\emph{2020
  IEEE International Symposium on High Performance Computer Architecture
  (HPCA)}}.
\newblock


\bibitem[\protect\citeauthoryear{Choquette, Gandhi, Giroux, Stam, and
  Krashinsky}{Choquette et~al\mbox{.}}{2021}]%
        {a100}
\bibfield{author}{\bibinfo{person}{Jack Choquette}, \bibinfo{person}{Wishwesh
  Gandhi}, \bibinfo{person}{Olivier Giroux}, \bibinfo{person}{Nick Stam}, {and}
  \bibinfo{person}{Ronny Krashinsky}.} \bibinfo{year}{2021}\natexlab{}.
\newblock \showarticletitle{Nvidia a100 tensor core gpu: Performance and
  innovation}.
\newblock \bibinfo{journal}{\emph{IEEE Micro}} \bibinfo{volume}{41},
  \bibinfo{number}{2} (\bibinfo{year}{2021}), \bibinfo{pages}{29--35}.
\newblock


\bibitem[\protect\citeauthoryear{Coorporation}{Coorporation}{2017}]%
        {v100}
\bibfield{author}{\bibinfo{person}{NVIDIA Coorporation}.}
  \bibinfo{year}{2017}\natexlab{}.
\newblock \bibinfo{booktitle}{\emph{NVIDIA Tesla V100 GPU Architecture}}.
\newblock \bibinfo{type}{{T}echnical {R}eport}.
\newblock
\newblock
\shownote{\url{http://www.nvidia.com/object/volta-architecture}.}


\bibitem[\protect\citeauthoryear{Dhakal, Cho, Kulkarni, Ramakrishnan, and
  Sharma}{Dhakal et~al\mbox{.}}{2020a}]%
        {mt_tvm}
\bibfield{author}{\bibinfo{person}{Aditya Dhakal}, \bibinfo{person}{Junguk
  Cho}, \bibinfo{person}{Sameer~G Kulkarni}, \bibinfo{person}{KK Ramakrishnan},
  {and} \bibinfo{person}{Puneet Sharma}.} \bibinfo{year}{2020}\natexlab{a}.
\newblock \showarticletitle{Spatial Sharing of GPU for Autotuning DNN models}.
\newblock \bibinfo{journal}{\emph{arXiv preprint arXiv:2008.03602}}
  (\bibinfo{year}{2020}).
\newblock


\bibitem[\protect\citeauthoryear{Dhakal, Kulkarni, and Ramakrishnan}{Dhakal
  et~al\mbox{.}}{2020b}]%
        {gslice}
\bibfield{author}{\bibinfo{person}{Aditya Dhakal}, \bibinfo{person}{Sameer~G
  Kulkarni}, {and} \bibinfo{person}{KK Ramakrishnan}.}
  \bibinfo{year}{2020}\natexlab{b}.
\newblock \showarticletitle{Gslice: controlled spatial sharing of gpus for a
  scalable inference platform}. In \bibinfo{booktitle}{\emph{Proceedings of the
  11th ACM Symposium on Cloud Computing}}. \bibinfo{pages}{492--506}.
\newblock


\bibitem[\protect\citeauthoryear{Ding, Zhu, Jia, Pekhimenko, and Han}{Ding
  et~al\mbox{.}}{2020}]%
        {ios}
\bibfield{author}{\bibinfo{person}{Yaoyao Ding}, \bibinfo{person}{Ligeng Zhu},
  \bibinfo{person}{Zhihao Jia}, \bibinfo{person}{Gennady Pekhimenko}, {and}
  \bibinfo{person}{Song Han}.} \bibinfo{year}{2020}\natexlab{}.
\newblock \showarticletitle{IOS: Inter-Operator Scheduler for CNN
  Acceleration}.
\newblock \bibinfo{journal}{\emph{arXiv preprint arXiv:2011.01302}}
  (\bibinfo{year}{2020}).
\newblock


\bibitem[\protect\citeauthoryear{Dumoulin and Visin}{Dumoulin and
  Visin}{2016}]%
        {direct_conv}
\bibfield{author}{\bibinfo{person}{Vincent Dumoulin} {and}
  \bibinfo{person}{Francesco Visin}.} \bibinfo{year}{2016}\natexlab{}.
\newblock \showarticletitle{A guide to convolution arithmetic for deep
  learning}.
\newblock \bibinfo{journal}{\emph{arXiv preprint arXiv:1603.07285}}
  (\bibinfo{year}{2016}).
\newblock


\bibitem[\protect\citeauthoryear{Fernandez}{Fernandez}{2022}]%
        {metaverse}
\bibfield{author}{\bibinfo{person}{Peter Fernandez}.}
  \bibinfo{year}{2022}\natexlab{}.
\newblock \showarticletitle{Facebook, Meta, the metaverse and libraries}.
\newblock \bibinfo{journal}{\emph{Library Hi Tech News}}
  (\bibinfo{year}{2022}).
\newblock


\bibitem[\protect\citeauthoryear{Ghodrati, Ahn, Kim, Kinzer, Yatham, Alla,
  Sharma, Alian, Ebrahimi, Kim, et~al\mbox{.}}{Ghodrati et~al\mbox{.}}{2020}]%
        {plana}
\bibfield{author}{\bibinfo{person}{Soroush Ghodrati},
  \bibinfo{person}{Byung~Hoon Ahn}, \bibinfo{person}{Joon~Kyung Kim},
  \bibinfo{person}{Sean Kinzer}, \bibinfo{person}{Brahmendra~Reddy Yatham},
  \bibinfo{person}{Navateja Alla}, \bibinfo{person}{Hardik Sharma},
  \bibinfo{person}{Mohammad Alian}, \bibinfo{person}{Eiman Ebrahimi},
  \bibinfo{person}{Nam~Sung Kim}, {et~al\mbox{.}}}
  \bibinfo{year}{2020}\natexlab{}.
\newblock \showarticletitle{Planaria: Dynamic architecture fission for spatial
  multi-tenant acceleration of deep neural networks}. In
  \bibinfo{booktitle}{\emph{2020 53rd Annual IEEE/ACM International Symposium
  on Microarchitecture (MICRO)}}. IEEE, \bibinfo{pages}{681--697}.
\newblock


\bibitem[\protect\citeauthoryear{He, Zhang, Ren, and Sun}{He
  et~al\mbox{.}}{2016}]%
        {r50}
\bibfield{author}{\bibinfo{person}{Kaiming He}, \bibinfo{person}{Xiangyu
  Zhang}, \bibinfo{person}{Shaoqing Ren}, {and} \bibinfo{person}{Jian Sun}.}
  \bibinfo{year}{2016}\natexlab{}.
\newblock \showarticletitle{Deep residual learning for image recognition}. In
  \bibinfo{booktitle}{\emph{Proceedings of the IEEE conference on computer
  vision and pattern recognition}}. \bibinfo{pages}{770--778}.
\newblock


\bibitem[\protect\citeauthoryear{Jia, Padon, Thomas, Warszawski, Zaharia, and
  Aiken}{Jia et~al\mbox{.}}{2019}]%
        {taso}
\bibfield{author}{\bibinfo{person}{Zhihao Jia}, \bibinfo{person}{Oded Padon},
  \bibinfo{person}{James Thomas}, \bibinfo{person}{Todd Warszawski},
  \bibinfo{person}{Matei Zaharia}, {and} \bibinfo{person}{Alex Aiken}.}
  \bibinfo{year}{2019}\natexlab{}.
\newblock \showarticletitle{TASO: optimizing deep learning computation with
  automatic generation of graph substitutions}. In
  \bibinfo{booktitle}{\emph{Proceedings of the 27th ACM Symposium on Operating
  Systems Principles}}. \bibinfo{pages}{47--62}.
\newblock


\bibitem[\protect\citeauthoryear{Lavin and Gray}{Lavin and Gray}{2016}]%
        {winograd}
\bibfield{author}{\bibinfo{person}{Andrew Lavin} {and} \bibinfo{person}{Scott
  Gray}.} \bibinfo{year}{2016}\natexlab{}.
\newblock \showarticletitle{Fast algorithms for convolutional neural networks}.
  In \bibinfo{booktitle}{\emph{Proceedings of the IEEE Conference on Computer
  Vision and Pattern Recognition}}. \bibinfo{pages}{4013--4021}.
\newblock


\bibitem[\protect\citeauthoryear{Li, Zheng, Pekhimenko, and Long}{Li
  et~al\mbox{.}}{2020}]%
        {horizon}
\bibfield{author}{\bibinfo{person}{Ao Li}, \bibinfo{person}{Bojian Zheng},
  \bibinfo{person}{Gennady Pekhimenko}, {and} \bibinfo{person}{Fan Long}.}
  \bibinfo{year}{2020}\natexlab{}.
\newblock \showarticletitle{Automatic horizontal fusion for GPU kernels}.
\newblock \bibinfo{journal}{\emph{arXiv preprint arXiv:2007.01277}}
  (\bibinfo{year}{2020}).
\newblock


\bibitem[\protect\citeauthoryear{Lui, Yetim, {\"O}zkan, Zhao, Tsai, Wu, and
  Hempstead}{Lui et~al\mbox{.}}{2021}]%
        {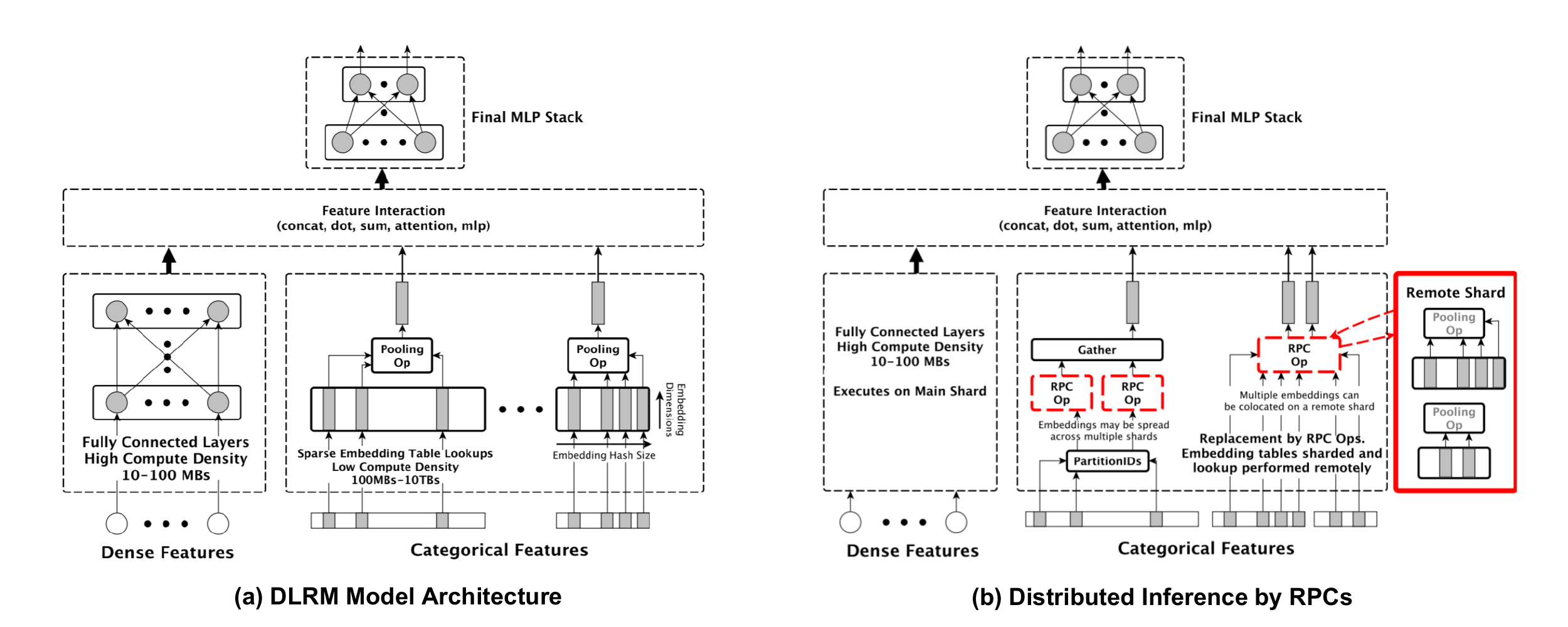}
\bibfield{author}{\bibinfo{person}{Michael Lui}, \bibinfo{person}{Yavuz Yetim},
  \bibinfo{person}{{\"O}zg{\"u}r {\"O}zkan}, \bibinfo{person}{Zhuoran Zhao},
  \bibinfo{person}{Shin-Yeh Tsai}, \bibinfo{person}{Carole-Jean Wu}, {and}
  \bibinfo{person}{Mark Hempstead}.} \bibinfo{year}{2021}\natexlab{}.
\newblock \showarticletitle{Understanding capacity-driven scale-out neural
  recommendation inference}. In \bibinfo{booktitle}{\emph{2021 IEEE
  International Symposium on Performance Analysis of Systems and Software
  (ISPASS)}}. IEEE, \bibinfo{pages}{162--171}.
\newblock


\bibitem[\protect\citeauthoryear{Mendoza, Romero, Li, Yadwadkar, and
  Kozyrakis}{Mendoza et~al\mbox{.}}{2021}]%
        {stanf}
\bibfield{author}{\bibinfo{person}{Daniel Mendoza}, \bibinfo{person}{Francisco
  Romero}, \bibinfo{person}{Qian Li}, \bibinfo{person}{Neeraja~J Yadwadkar},
  {and} \bibinfo{person}{Christos Kozyrakis}.} \bibinfo{year}{2021}\natexlab{}.
\newblock \showarticletitle{Interference-Aware Scheduling for Inference
  Serving}. In \bibinfo{booktitle}{\emph{Proceedings of the 1st Workshop on
  Machine Learning and Systems}}. \bibinfo{pages}{80--88}.
\newblock


\bibitem[\protect\citeauthoryear{Niu, Guan, Wang, Agrawal, and Ren}{Niu
  et~al\mbox{.}}{2021}]%
        {advanced}
\bibfield{author}{\bibinfo{person}{Wei Niu}, \bibinfo{person}{Jiexiong Guan},
  \bibinfo{person}{Yanzhi Wang}, \bibinfo{person}{Gagan Agrawal}, {and}
  \bibinfo{person}{Bin Ren}.} \bibinfo{year}{2021}\natexlab{}.
\newblock \showarticletitle{DNNFusion: accelerating deep neural networks
  execution with advanced operator fusion}. In
  \bibinfo{booktitle}{\emph{Proceedings of the 42nd ACM SIGPLAN International
  Conference on Programming Language Design and Implementation}}.
  \bibinfo{pages}{883--898}.
\newblock


\bibitem[\protect\citeauthoryear{NVIDIA}{NVIDIA}{2013}]%
        {hyperq}
\bibfield{author}{\bibinfo{person}{NVIDIA}.} \bibinfo{year}{2013}\natexlab{}.
\newblock \bibinfo{title}{Hyper-Q}.
\newblock
\newblock
\newblock
\shownote{\url{https://developer.download.nvidia.com/compute/DevZone/C/html_x64/6_Advanced/simpleHyperQ/doc/HyperQ.pdf}.}


\bibitem[\protect\citeauthoryear{NVIDIA}{NVIDIA}{2015}]%
        {stream}
\bibfield{author}{\bibinfo{person}{NVIDIA}.} \bibinfo{year}{2015}\natexlab{}.
\newblock \bibinfo{title}{CUDA Multi-Streams}.
\newblock
\newblock
\newblock
\shownote{\url{https://developer.nvidia.com/blog/gpu-pro-tip-cuda-7-streams-simplify-concurrency/}.}


\bibitem[\protect\citeauthoryear{NVIDIA}{NVIDIA}{2020a}]%
        {cuda}
\bibfield{author}{\bibinfo{person}{NVIDIA}.} \bibinfo{year}{2020}\natexlab{a}.
\newblock \bibinfo{title}{CUDA Programming Guide}.
\newblock
\newblock
\newblock
\shownote{\url{https://docs.nvidia.com/cuda/cuda-c-programming-guide/index.html}.}


\bibitem[\protect\citeauthoryear{NVIDIA}{NVIDIA}{2020b}]%
        {mig}
\bibfield{author}{\bibinfo{person}{NVIDIA}.} \bibinfo{year}{2020}\natexlab{b}.
\newblock \bibinfo{title}{NVIDIA Multi Instance GPU (MIG)}.
\newblock
\newblock
\newblock
\shownote{\url{https://docs.nvidia.com/datacenter/tesla/mig-user-guide/}.}


\bibitem[\protect\citeauthoryear{NVIDIA}{NVIDIA}{2020c}]%
        {mps}
\bibfield{author}{\bibinfo{person}{NVIDIA}.} \bibinfo{year}{2020}\natexlab{c}.
\newblock \bibinfo{title}{NVIDIA Multi Process Service (MPS)}.
\newblock
\newblock
\newblock
\shownote{\url{https://docs.nvidia.com/deploy/pdf/CUDA-Multi-Process-Service-Overview.pdf}.}


\bibitem[\protect\citeauthoryear{NVIDIA}{NVIDIA}{2020d}]%
        {vcs}
\bibfield{author}{\bibinfo{person}{NVIDIA}.} \bibinfo{year}{2020}\natexlab{d}.
\newblock \bibinfo{title}{NVIDIA Virtual Compute Server}.
\newblock
\newblock
\newblock
\shownote{\url{https://www.nvidia.com/content/dam/en-zz/Solutions/design-visualization/solutions/resources/documents1/Technical-Brief-Multi-Instance-GPU-NVIDIA-Virtual-Compute-Server.pdf}.}


\bibitem[\protect\citeauthoryear{NVIDIA}{NVIDIA}{2021}]%
        {graph}
\bibfield{author}{\bibinfo{person}{NVIDIA}.} \bibinfo{year}{2021}\natexlab{}.
\newblock \bibinfo{title}{CUDA Graphs}.
\newblock
\newblock
\newblock
\shownote{\url{https://developer.nvidia.com/blog/cuda-graphs/}.}


\bibitem[\protect\citeauthoryear{Parisi, Kemker, Part, Kanan, and
  Wermter}{Parisi et~al\mbox{.}}{2019}]%
        {lifelong}
\bibfield{author}{\bibinfo{person}{German~I Parisi}, \bibinfo{person}{Ronald
  Kemker}, \bibinfo{person}{Jose~L Part}, \bibinfo{person}{Christopher Kanan},
  {and} \bibinfo{person}{Stefan Wermter}.} \bibinfo{year}{2019}\natexlab{}.
\newblock \showarticletitle{Continual lifelong learning with neural networks: A
  review}.
\newblock \bibinfo{journal}{\emph{Neural Networks}}  \bibinfo{volume}{113}
  (\bibinfo{year}{2019}), \bibinfo{pages}{54--71}.
\newblock


\bibitem[\protect\citeauthoryear{Qin, Yu, Liu, and Chen}{Qin
  et~al\mbox{.}}{2018}]%
        {filter_prune}
\bibfield{author}{\bibinfo{person}{Zhuwei Qin}, \bibinfo{person}{Fuxun Yu},
  \bibinfo{person}{Chenchen Liu}, {and} \bibinfo{person}{Xiang Chen}.}
  \bibinfo{year}{2018}\natexlab{}.
\newblock \showarticletitle{Functionality-oriented convolutional filter
  pruning}.
\newblock \bibinfo{journal}{\emph{arXiv preprint arXiv:1810.07322}}
  (\bibinfo{year}{2018}).
\newblock


\bibitem[\protect\citeauthoryear{Ren, Xiao, Chang, Huang, Li, Gupta, Chen, and
  Wang}{Ren et~al\mbox{.}}{2021}]%
        {active}
\bibfield{author}{\bibinfo{person}{Pengzhen Ren}, \bibinfo{person}{Yun Xiao},
  \bibinfo{person}{Xiaojun Chang}, \bibinfo{person}{Po-Yao Huang},
  \bibinfo{person}{Zhihui Li}, \bibinfo{person}{Brij~B Gupta},
  \bibinfo{person}{Xiaojiang Chen}, {and} \bibinfo{person}{Xin Wang}.}
  \bibinfo{year}{2021}\natexlab{}.
\newblock \showarticletitle{A survey of deep active learning}.
\newblock \bibinfo{journal}{\emph{ACM Computing Surveys (CSUR)}}
  \bibinfo{volume}{54}, \bibinfo{number}{9} (\bibinfo{year}{2021}),
  \bibinfo{pages}{1--40}.
\newblock


\bibitem[\protect\citeauthoryear{Reports}{Reports}{2021}]%
        {gpu_report}
\bibfield{author}{\bibinfo{person}{Market Reports}.}
  \bibinfo{year}{2021}\natexlab{}.
\newblock \bibinfo{title}{Global Data Center Accelerator Market Size, Status
  and Forecast 2020-2025}.
\newblock
\newblock
\newblock
\shownote{\url{https://www.mynewsdesk.com/brandessence/pressreleases/data-center-accelerator-market-size-2021-cagr-38-dot-7-percent-3112488}.}


\bibitem[\protect\citeauthoryear{Sampson, Chen, and Roesch}{Sampson
  et~al\mbox{.}}{2022}]%
        {tvm_talk}
\bibfield{author}{\bibinfo{person}{Adrian Sampson}, \bibinfo{person}{Tianqi
  Chen}, {and} \bibinfo{person}{Jared Roesch}.}
  \bibinfo{year}{2022}\natexlab{}.
\newblock \bibinfo{title}{Apache TVM Unity: a vision for the ML software and
  hardware ecosystem}.
\newblock
\newblock
\newblock
\shownote{\url{https://tvm.apache.org/2021/12/15/tvm-unity}.}


\bibitem[\protect\citeauthoryear{Shao, Clemons, Venkatesan, Zimmer, Fojtik,
  Jiang, Keller, Klinefelter, Pinckney, Raina, et~al\mbox{.}}{Shao
  et~al\mbox{.}}{2019}]%
        {chiplet2}
\bibfield{author}{\bibinfo{person}{Yakun~Sophia Shao}, \bibinfo{person}{Jason
  Clemons}, \bibinfo{person}{Rangharajan Venkatesan}, \bibinfo{person}{Brian
  Zimmer}, \bibinfo{person}{Matthew Fojtik}, \bibinfo{person}{Nan Jiang},
  \bibinfo{person}{Ben Keller}, \bibinfo{person}{Alicia Klinefelter},
  \bibinfo{person}{Nathaniel Pinckney}, \bibinfo{person}{Priyanka Raina},
  {et~al\mbox{.}}} \bibinfo{year}{2019}\natexlab{}.
\newblock \showarticletitle{Simba: Scaling deep-learning inference with
  multi-chip-module-based architecture}. In
  \bibinfo{booktitle}{\emph{Proceedings of the 52nd Annual IEEE/ACM
  International Symposium on Microarchitecture}}. \bibinfo{pages}{14--27}.
\newblock


\bibitem[\protect\citeauthoryear{Shen, Roesch, Chen, Chen, Wu, Li, Sharma,
  Tatlock, and Wang}{Shen et~al\mbox{.}}{2021}]%
        {nimble}
\bibfield{author}{\bibinfo{person}{Haichen Shen}, \bibinfo{person}{Jared
  Roesch}, \bibinfo{person}{Zhi Chen}, \bibinfo{person}{Wei Chen},
  \bibinfo{person}{Yong Wu}, \bibinfo{person}{Mu Li}, \bibinfo{person}{Vin
  Sharma}, \bibinfo{person}{Zachary Tatlock}, {and} \bibinfo{person}{Yida
  Wang}.} \bibinfo{year}{2021}\natexlab{}.
\newblock \showarticletitle{Nimble: Efficiently compiling dynamic neural
  networks for model inference}.
\newblock \bibinfo{journal}{\emph{Proceedings of MLSys}}
  (\bibinfo{year}{2021}).
\newblock


\bibitem[\protect\citeauthoryear{Smith, Patwary, Norick, LeGresley,
  Rajbhandari, Casper, Liu, Prabhumoye, Zerveas, Korthikanti,
  et~al\mbox{.}}{Smith et~al\mbox{.}}{2022}]%
        {mt_nlg}
\bibfield{author}{\bibinfo{person}{Shaden Smith}, \bibinfo{person}{Mostofa
  Patwary}, \bibinfo{person}{Brandon Norick}, \bibinfo{person}{Patrick
  LeGresley}, \bibinfo{person}{Samyam Rajbhandari}, \bibinfo{person}{Jared
  Casper}, \bibinfo{person}{Zhun Liu}, \bibinfo{person}{Shrimai Prabhumoye},
  \bibinfo{person}{George Zerveas}, \bibinfo{person}{Vijay Korthikanti},
  {et~al\mbox{.}}} \bibinfo{year}{2022}\natexlab{}.
\newblock \showarticletitle{Using DeepSpeed and Megatron to Train
  Megatron-Turing NLG 530B, A Large-Scale Generative Language Model}.
\newblock \bibinfo{journal}{\emph{arXiv preprint arXiv:2201.11990}}
  (\bibinfo{year}{2022}).
\newblock


\bibitem[\protect\citeauthoryear{Soifer, Li, Li, Zhu, Li, He, Zheng, Oltean,
  Mosyak, Barnes, et~al\mbox{.}}{Soifer et~al\mbox{.}}{2019}]%
        {dlis}
\bibfield{author}{\bibinfo{person}{Jonathan Soifer}, \bibinfo{person}{Jason
  Li}, \bibinfo{person}{Mingqin Li}, \bibinfo{person}{Jeffrey Zhu},
  \bibinfo{person}{Yingnan Li}, \bibinfo{person}{Yuxiong He},
  \bibinfo{person}{Elton Zheng}, \bibinfo{person}{Adi Oltean},
  \bibinfo{person}{Maya Mosyak}, \bibinfo{person}{Chris Barnes},
  {et~al\mbox{.}}} \bibinfo{year}{2019}\natexlab{}.
\newblock \showarticletitle{Deep learning inference service at microsoft}. In
  \bibinfo{booktitle}{\emph{2019 $\{$USENIX$\}$ Conference on Operational
  Machine Learning (OpML 19)}}. \bibinfo{pages}{15--17}.
\newblock


\bibitem[\protect\citeauthoryear{Sun, Agostini, Dong, and Kaeli}{Sun
  et~al\mbox{.}}{2019}]%
        {gpu_stat}
\bibfield{author}{\bibinfo{person}{Yifan Sun}, \bibinfo{person}{Nicolas~Bohm
  Agostini}, \bibinfo{person}{Shi Dong}, {and} \bibinfo{person}{David Kaeli}.}
  \bibinfo{year}{2019}\natexlab{}.
\newblock \showarticletitle{Summarizing CPU and GPU design trends with product
  data}.
\newblock \bibinfo{journal}{\emph{arXiv preprint arXiv:1911.11313}}
  (\bibinfo{year}{2019}).
\newblock


\bibitem[\protect\citeauthoryear{Tan, Li, and et~al.}{Tan
  et~al\mbox{.}}{2021}]%
        {mig_serving}
\bibfield{author}{\bibinfo{person}{Cheng Tan}, \bibinfo{person}{Zhichao Li},
  {and} \bibinfo{person}{et al.}} \bibinfo{year}{2021}\natexlab{}.
\newblock \showarticletitle{Serving DNN Models with Multi-Instance GPUs: A Case
  of the Reconfigurable Machine Scheduling Problem}.
\newblock \bibinfo{journal}{\emph{arXiv:2109.11067}} (\bibinfo{year}{2021}).
\newblock


\bibitem[\protect\citeauthoryear{TensorFlow}{TensorFlow}{2020}]%
        {xla}
\bibfield{author}{\bibinfo{person}{TensorFlow}.}
  \bibinfo{year}{2020}\natexlab{}.
\newblock \showarticletitle{TensorFlow XLA (Accelerated Linear Algebra)}.
\newblock
\newblock
\shownote{\url{https://www.tensorflow.org/xla}.}


\bibitem[\protect\citeauthoryear{Vanholder}{Vanholder}{2016}]%
        {trt}
\bibfield{author}{\bibinfo{person}{Han Vanholder}.}
  \bibinfo{year}{2016}\natexlab{}.
\newblock \showarticletitle{Efficient inference with tensorrt}. In
  \bibinfo{booktitle}{\emph{GPU Technology Conference}},
  Vol.~\bibinfo{volume}{1}. \bibinfo{pages}{2}.
\newblock


\bibitem[\protect\citeauthoryear{Wesolowski, Acun, Andrei, Aziz, Dankel, Gregg,
  Meng, Meurillon, Sheahan, Tian, et~al\mbox{.}}{Wesolowski
  et~al\mbox{.}}{2021}]%
        {fb_center}
\bibfield{author}{\bibinfo{person}{Lukasz Wesolowski}, \bibinfo{person}{Bilge
  Acun}, \bibinfo{person}{Valentin Andrei}, \bibinfo{person}{Adnan Aziz},
  \bibinfo{person}{Gisle Dankel}, \bibinfo{person}{Christopher Gregg},
  \bibinfo{person}{Xiaoqiao Meng}, \bibinfo{person}{Cyril Meurillon},
  \bibinfo{person}{Denis Sheahan}, \bibinfo{person}{Lei Tian}, {et~al\mbox{.}}}
  \bibinfo{year}{2021}\natexlab{}.
\newblock \showarticletitle{Datacenter-Scale Analysis and Optimization of GPU
  Machine Learning Workloads}.
\newblock \bibinfo{journal}{\emph{IEEE Micro}} \bibinfo{volume}{41},
  \bibinfo{number}{5} (\bibinfo{year}{2021}), \bibinfo{pages}{101--112}.
\newblock


\bibitem[\protect\citeauthoryear{Wu, Xu, and Wang}{Wu et~al\mbox{.}}{2020}]%
        {irina}
\bibfield{author}{\bibinfo{person}{Xiaorui Wu}, \bibinfo{person}{Hong Xu},
  {and} \bibinfo{person}{Yi Wang}.} \bibinfo{year}{2020}\natexlab{}.
\newblock \showarticletitle{Irina: Accelerating DNN Inference with Efficient
  Online Scheduling}. In \bibinfo{booktitle}{\emph{4th Asia-Pacific Workshop on
  Networking}}. \bibinfo{pages}{36--43}.
\newblock


\bibitem[\protect\citeauthoryear{Yang, Phothilimthana, Wang, Willsey, Roy, and
  Pienaar}{Yang et~al\mbox{.}}{2021}]%
        {mlsys}
\bibfield{author}{\bibinfo{person}{Yichen Yang}, \bibinfo{person}{Phitchaya
  Phothilimthana}, \bibinfo{person}{Yisu Wang}, \bibinfo{person}{Max Willsey},
  \bibinfo{person}{Sudip Roy}, {and} \bibinfo{person}{Jacques Pienaar}.}
  \bibinfo{year}{2021}\natexlab{}.
\newblock \showarticletitle{Equality saturation for tensor graph
  superoptimization}.
\newblock \bibinfo{journal}{\emph{Proceedings of Machine Learning and Systems}}
   \bibinfo{volume}{3} (\bibinfo{year}{2021}), \bibinfo{pages}{255--268}.
\newblock


\bibitem[\protect\citeauthoryear{Yu and et~al.}{Yu and et~al.}{2021}]%
        {iccad}
\bibfield{author}{\bibinfo{person}{Fuxun Yu} {and} \bibinfo{person}{et al.}}
  \bibinfo{year}{2021}\natexlab{}.
\newblock \showarticletitle{Automated Runtime-Aware Scheduling for Multi-Tenant
  DNN Inference on GPU}. In \bibinfo{booktitle}{\emph{Proceedings of the 40th
  IEEE International Conference on Computer Aided Design (ICCAD)}}.
\newblock


\bibitem[\protect\citeauthoryear{Yu, Liu, Wang, Wang, and Chen}{Yu
  et~al\mbox{.}}{2020a}]%
        {feature_prune}
\bibfield{author}{\bibinfo{person}{Fuxun Yu}, \bibinfo{person}{Chenchen Liu},
  \bibinfo{person}{Di Wang}, \bibinfo{person}{Yanzhi Wang}, {and}
  \bibinfo{person}{Xiang Chen}.} \bibinfo{year}{2020}\natexlab{a}.
\newblock \showarticletitle{AntiDote: Attention-based Dynamic Optimization for
  Neural Network Runtime Efficiency}. In \bibinfo{booktitle}{\emph{2020 Design,
  Automation \& Test in Europe Conference \& Exhibition (DATE)}}.
\newblock


\bibitem[\protect\citeauthoryear{Yu, Qin, Wang, Xu, Liu, Tian, and Chen}{Yu
  et~al\mbox{.}}{2020b}]%
        {connect_prune}
\bibfield{author}{\bibinfo{person}{Fuxun Yu}, \bibinfo{person}{Zhuwei Qin},
  \bibinfo{person}{Di Wang}, \bibinfo{person}{Ping Xu},
  \bibinfo{person}{Chenchen Liu}, \bibinfo{person}{Zhi Tian}, {and}
  \bibinfo{person}{Xiang Chen}.} \bibinfo{year}{2020}\natexlab{b}.
\newblock \showarticletitle{Dc-cnn: computational flow redefinition for
  efficient cnn through structural decoupling}. In
  \bibinfo{booktitle}{\emph{2020 Design, Automation \& Test in Europe
  Conference \& Exhibition (DATE)}}. IEEE, \bibinfo{pages}{1097--1102}.
\newblock


\bibitem[\protect\citeauthoryear{Zhao, Barijough, and Gerstlauer}{Zhao
  et~al\mbox{.}}{2018}]%
        {deepthing}
\bibfield{author}{\bibinfo{person}{Zhuoran Zhao},
  \bibinfo{person}{Kamyar~Mirzazad Barijough}, {and} \bibinfo{person}{Andreas
  Gerstlauer}.} \bibinfo{year}{2018}\natexlab{}.
\newblock \showarticletitle{Deepthings: Distributed adaptive deep learning
  inference on resource-constrained iot edge clusters}.
\newblock \bibinfo{journal}{\emph{IEEE Transactions on Computer-Aided Design of
  Integrated Circuits and Systems}} \bibinfo{volume}{37}, \bibinfo{number}{11}
  (\bibinfo{year}{2018}), \bibinfo{pages}{2348--2359}.
\newblock


\bibitem[\protect\citeauthoryear{Zheng, Wang, and Louri}{Zheng
  et~al\mbox{.}}{2020b}]%
        {chiplet1}
\bibfield{author}{\bibinfo{person}{Hao Zheng}, \bibinfo{person}{Ke Wang}, {and}
  \bibinfo{person}{Ahmed Louri}.} \bibinfo{year}{2020}\natexlab{b}.
\newblock \showarticletitle{A versatile and flexible chiplet-based system
  design for heterogeneous manycore architectures}. In
  \bibinfo{booktitle}{\emph{2020 57th ACM/IEEE Design Automation Conference
  (DAC)}}. IEEE, \bibinfo{pages}{1--6}.
\newblock


\bibitem[\protect\citeauthoryear{Zheng, Jia, Sun, Wu, Yu, Haj-Ali, Wang, Yang,
  Zhuo, Sen, et~al\mbox{.}}{Zheng et~al\mbox{.}}{2020a}]%
        {ansor}
\bibfield{author}{\bibinfo{person}{Lianmin Zheng}, \bibinfo{person}{Chengfan
  Jia}, \bibinfo{person}{Minmin Sun}, \bibinfo{person}{Zhao Wu},
  \bibinfo{person}{Cody~Hao Yu}, \bibinfo{person}{Ameer Haj-Ali},
  \bibinfo{person}{Yida Wang}, \bibinfo{person}{Jun Yang},
  \bibinfo{person}{Danyang Zhuo}, \bibinfo{person}{Koushik Sen},
  {et~al\mbox{.}}} \bibinfo{year}{2020}\natexlab{a}.
\newblock \showarticletitle{Ansor: Generating high-performance tensor programs
  for deep learning}. In \bibinfo{booktitle}{\emph{14th $\{$USENIX$\}$
  Symposium on Operating Systems Design and Implementation ($\{$OSDI$\}$ 20)}}.
  \bibinfo{pages}{863--879}.
\newblock


\bibitem[\protect\citeauthoryear{Zhou, Wen, Teodorescu, and Du}{Zhou
  et~al\mbox{.}}{2019}]%
        {hotedge}
\bibfield{author}{\bibinfo{person}{Li Zhou}, \bibinfo{person}{Hao Wen},
  \bibinfo{person}{Radu Teodorescu}, {and} \bibinfo{person}{David~HC Du}.}
  \bibinfo{year}{2019}\natexlab{}.
\newblock \showarticletitle{Distributing deep neural networks with
  containerized partitions at the edge}. In \bibinfo{booktitle}{\emph{2nd
  $\{$USENIX$\}$ Workshop on Hot Topics in Edge Computing (HotEdge 19)}}.
\newblock


\end{thebibliography}
